\documentclass[a4paper,aps,pra,twocolumn,preprintnumbers,amsmath,amssymb]{
revtex4}
\usepackage{amsmath}
\usepackage{verbatim}
\usepackage{graphicx}

 \newcommand{\ket}[1]{\left|#1\right\rangle} 
 \newcommand{\bra}[1]{\left\langle#1\right|} 
 \newcommand{\braket}[2]{\left< #1 \vphantom{#2} \right|
  \left. #2 \vphantom{#1} \right>} 

\begin{document}
 
\title{Quench dynamics of dipolar fermions in a one-dimensional harmonic trap}

\author{ Tobias Gra\ss}
\affiliation{ICFO-Institut de Ci\`encies Fot\`oniques, Av. Carl Friedrich
Gauss 3, 08860 Barcelona, Spain}

\begin{abstract}
We study a system of few fermions in a one-dimensional harmonic trap, and focus on the case of dipolar majority particles in contact with a single impurity. The impurity is used both for quenching the system, and for tracking the system evolution after the quench. Employing exact diagonalization, we investigate relaxation and thermalization properties. In the absence of dipolar interactions, the system is near integrability, and the dynamics remains oscillatory even on long time scales. On the other hand, repulsive as well as attractive dipolar interactions lead to quick relaxation to the diagonal ensemble average which is significantly different from corresponding thermal averages. A Wigner-shaped level spacing distribution indicates level repulsion and thus chaotic dynamical behavior due to the presence of dipolar interactions.
\end{abstract}

\maketitle

\section{Introduction}
From the microscopic point of view physical systems are governed by the laws of quantum mechanics. For systems of macroscopic size, however, a description on this level usually fails due to the huge amount of different degrees of freedom. In such case, a tractable approach is to take appropriate averages over the degrees of freedom, replacing the microscopic laws by the laws of statistical mechanics. This switch has some paradoxical consequences, such as the emergence of irreversible dynamics from microscopic laws with time-reversal symmetry. 
In order to reconcile the physical laws on the microscopic and the macroscopic level, it seems to be necessary to extend our microscopic understanding towards larger system sizes. Therefore, we need to improve our capabilities in obtaining exact or quasi-exact knowledge about a system. 

Maybe the most promising direction to achieve this goal are quantum simulations, that is, experiments with well-controlled, tunable quantum systems \cite{mlbook}. A remarkable step towards bridging the gap between microscopic and macroscopic physics were the recent experiments in Heidelberg with cold fermionic atoms in a one-dimensional trap \cite{jochim11,jochim12,jochim13}. Outstanding through the precise control over particle numbers, the experiment of Ref. \cite{jochim13} demonstrated the formation of a Fermi ``sea'' of up to five majority atoms interacting with a single impurity atom. These experimental advances have also boosted the theoretical interest in one-dimensionally trapped Fermi systems \cite{brouzos13,sowinski13,astrakharchik13,blume13,zinner14,ho14,blume14,valiente14,mehta14,miguelangel14,blume15,levinsen,QChem}.

In this article, we extend the scenario from the Heidelberg experiments into two directions: 
First, we assume dipolar interactions between the fermions. 
Second, we shift the focus onto dynamical processes.
 The ``standard'' interactions in quantum gases are contact interactions. With this, a spin-polarized Fermi sea as in Ref. \cite{jochim13} remains non-interacting, since contact interactions are forbidden by the Pauli principle. Clearly, the most prominent fermions in nature are electrons with long-range Coulomb interactions. To design an atomic quantum simulation of a Fermi systems with long-range repulsion, one might stick to dipolar interactions. The long-range character of these interactions is certainly one of various reasons for the huge research interest in dipolar quantum gases \cite{maciek-dipole2000,maciek-dipole2003,maciek-dipoleOL,pfau07,aristeu10,grass11,baranov-review,ferlaino,laburthe15}. With advances in cooling molecular systems \cite{jin08,debbie13}, and recent developments involving Rydberg gases \cite{buechler08,pfau12,adams15}, new experimental platforms involving strong dipolar interactions are emerging. One-dimensional systems with dipolar interactions can provide a particularly interesting scenario, as the confinement modifies the effective interaction potential with the possibility of confinement-induced resonances \cite{sinha-santos07,deuretzbacher10,guan14}.

In a system which is strongly confined to one dimension, dipolar interactions become almost local, and therefore barely modify the ground state of a small Fermi gas. Nevertheless, the dynamical behavior of the system might be affected strongly. Accordingly, our paper is concerned with the dynamics of the system. Certainly, from the point of view of numerics, dynamics is the most difficult aspect of a quantum system, and therefore best-suited to be studied via a quantum simulation. Quantum systems out-of-equilibrium, and the relation between unitary time evolution and thermodynamics, has attracted broad research interest, cf. Refs. \cite{schmiedmayer07,barthel08,cramer08,rigol09,rigol13,steinigeweg14,wouter14,huse14} or the review articles \cite{polkovnikov,gogolin-review}. A closely related subject are Lieb-Robinson bounds \cite{lieb-robinson} and the velocity of correlation spreading, in particular in the presence of long-range interactions \cite{hastings06,luca-philipp,carleo14,epj,jurcevic,richerme,carleo15}.

In this paper, we will study a one-dimensional Fermi system in which majority atoms interact, via contact, with a single impurity, as in the Heidelberg experiments. Among themselves, the majority atoms can interact via dipolar interactions. We will first provide a brief description of this system, and introduce the effective dipole-dipole interaction in Section \ref{Sec:System}. In Section \ref{Sec:III}, we will sketch the concept of time evolution in quantum mechanics, and define some useful quantities to describe the dynamics of a many-body system. After briefly studying static properties of the ground state in Section \ref{Sec:Stat}, we will focus on relaxation processes from an out-of-equilibrium state in Section \ref{Sec:Dyn}. Here, the impurity will serve, in a two-fold way, as a control knob: On the one hand, we can excite the system by exciting the impurity atom to higher oscillator levels. On the other hand, the impurity can serve as a ``thermometer''. We will investigate under which conditions the energy of the impurity atom relaxes to a constant value, and whether this value corresponds to the thermalized value. It will turn out that dipolar interactions between the majority atoms significantly enhance relaxation. Generally, the diagonal ensemble describes well the relaxed state, but it differs from thermalized values. Finally, in Section \ref{Discussion}, we will relate our results to more general properties of the system and analyze the level spacing distribution. Although not an integrable model, the system without dipolar interactions exhibits a strong tendency of level clustering. On the other hand, a Wigner-shaped distribution for the system with dipolar interactions indicates level repulsion and chaotic dynamics. In this sense, the level spacing distribution turns out to be a good indicator for the system's relaxation properties. In the appendix, we calculate the interaction matrix elements of a dipolar gas in one dimension.

\section{System}
\label{Sec:System}
We consider a system of fermionic atoms in an effectively one-dimensional harmonic trap $V_{\rm trap} = \frac{m}{2}\omega^2 x^2$, where $m$ is the mass of the atoms, and $\omega$ the axial trapping frequency, chosen along the $x$-direction. The one-dimensional trapping is achieved via a sufficiently strong anisotropy of the three-dimensional trap, freezing out the dynamics in the transverse directions. The ratio between axial trapping frequency $\omega$ and the transverse trapping frequency $\omega_{\perp}$ is given by the parameter $\lambda = \sqrt{\omega/\omega_{\perp}} = l_\perp/l$, with $l=\sqrt{\hbar/(m\omega)}$ and $l_\perp=\sqrt{\hbar/(m\omega_\perp)}$ the corresponding harmonic oscillator lengths. For most of our numerical studies, we choose $\lambda=1/3$, the value reported in Refs. \cite{jochim11,jochim13}. We will work in harmonic oscillator units of the axial trapping, that is, $\omega=1$, $l=1$, $m=1$, and $\hbar=1$.

The fermions can be prepared in different internal states, or, equivalently, the system may consist of different species with equal mass. We restrict ourselves to two-component systems, where $N_\uparrow$ and $N_\downarrow$ denotes the number of fermions in each component. We focus on the case where several majority atoms (forming the $\uparrow$-component) interact with a single impurity atom ($N_\downarrow=1$) via contact interactions, $V_{\rm contact} = g \sum_{i\in \uparrow, j \in \downarrow} \delta(x_i-x_j)$ with effective interaction strength $g$. Such scenario has been studied experimentally in Ref. \cite{jochim13}. Since identical fermions do not interact with each other via contact interaction, the Fermi sea of the majority systems would be non-interacting. To overcome this, we may equip the majority atoms with an electric or magnetic dipole moment $d$, which leads to dipole-dipole interactions between the majority atoms. For simplicity, we assume that the impurity atom carries no dipole moment, such that it interacts with the majority atoms only via contact interaction.

The dipoles shall be aligned, forming an angle $\theta$ with the $x$-axis. Then, the scattering between dipoles, separated by a vector $\vec{r}$, is described by the usual interaction potential $V_{\rm dd}(\vec{r}) = \frac{d^2}{r^3} (1-3 \cos^2\theta_{rd})$, with $\theta_{rd}$ the angle between the dipole and $\vec{r}$. Due to the strong transverse confinement, one can assume that transverse degrees of freedom will not be excited, and integrate out the $y$ and $z$ variables. This leads to the effective potential \cite{sinha-santos07,deuretzbacher10,guan14}
\begin{align}
V_{\rm dd}^{\rm eff}(x) = - \frac{d^2[1+3\cos(2\theta)]}{8\lambda l} U(x/\lambda) \equiv u U(x/\lambda),   
\end{align}
with the dimensionless potential
\begin{align}
 U(x)=-2|x|+\sqrt{2\pi}(1+x^2)e^{x^2/2}{\rm erfc}(x/\sqrt{2}),
\end{align}
where ${\rm erfc}$ is the complementary error function. An additional term $\propto \delta(x)$ occurring in the effective potential can be neglected for identical fermions. 
We note that the strength of the dipolar interactions, $u$, can be tuned via the polarization angle $\theta$, and can be made attractive or repulsive.

The total Hamiltonian will be studied by exact diagonalization in the Fock basis of the harmonic oscillator levels. We truncate this basis, taking into account up to 22 single-particle levels, for systems of up to five majority atoms. We consider interaction strengths up to $g=10$ and $|u|=10$. In second quantization, the Hamiltonian reads (assuming no dipole moment for the impurity $\downarrow$ atoms):
\begin{align}
\label{H}
 H=&  \sum_{j} j a_j^\dagger a_j + \frac{g}{2} \sum_{(ik) \in \uparrow,(jl) \in \downarrow} V_{ijkl}^{\rm contact} a_i^\dagger a_j^\dagger a_k a_l 
 \nonumber \\
& + \frac{u}{2} \sum_{(ijkl) \in \uparrow} V_{ijkl}^{\rm dipolar} a_i^\dagger a_j^\dagger a_k a_l,  
\end{align}
with annihilation/creation operator $a_j$ and $a_j^\dagger$, where the index $j$ refers to the $j$th orbital, while the index for the spin has been suppressed. A constant energy term $N\hbar \omega/2$, corresponding to the zero-point energy of the harmonic oscillator with $N$ particles, is neglected. With $\varphi_j(x)$ denoting the normalized orbitals, the interaction matrix elements are defined as follows:
\begin{align}
 V_{ijkl}^{\rm contact} =& \int_{-\infty}^{\infty} dx \ dx' \varphi_i(x) \varphi_j(x') \varphi_k(x) \varphi_l(x')
\\ \nonumber &\hspace{-2.2cm} {\rm and}& \\
 V_{ijkl}^{\rm dipolar} = & \int_{-\infty}^{\infty} dx \ dx' \ U(x/\lambda-x'/\lambda)  \nonumber \\
   & \times \varphi_i(x) \varphi_j(x') \varphi_k(x) \varphi_l(x').
\end{align}
Both integrals can be solved by integrating out the center-of-mass coordinate $x+x'$. While the remaining integral in the relative coordinate $x-x'$ is trivial for the contact interaction, some care must be taken for the dipolar potential. We describe how to evaluate this integral in the appendix.

\section{Time evolution and thermodynamics}
\label{Sec:III}
In this section, we will define different quantities which are useful for analyzing the dynamical properties of a quantum system. We will be interested in scenarios where the system is initially prepared in some pure state $\ket{\Psi_{\rm ini}}$, and then evolves in the Schr{\"o}dinger picture under the time-independent Hamiltonian $H$. The time-evolved state is then given by
\begin{align}
\label{Psi}
 \ket{\Psi(t)} = e^{-i H t} \ket{\Psi_{\rm ini}} = \sum_\alpha c_\alpha e^{-i E_\alpha t} \ket{\alpha}.
\end{align}
Here, we have decomposed the initial state into eigenstates $\ket{\alpha}$ of $H$, $H \ket{\alpha} = E_\alpha \ket{\alpha}$. The coefficients $c_\alpha$ are given by $c_\alpha = \braket{\alpha}{\Psi_{\rm ini}}$.

Using exact diagonalization, we can determine the full energy spectrum of a small system, and then straightforwardly calculate the evolution of an arbitrary state. The state vector itself, however, is not accessible experimentally. In practice, the system dynamics can be tracked by looking at the evolution of any observable. From the computational point of view, this requires to evaluate the quantum average of operators $O$: $\langle O(t) \rangle = \bra{\Psi(t)}O\ket{\Psi(t)}$. If all eigenstates $\alpha$ are non-degenerate, $\langle O(t) \rangle$ will, for sufficiently long times, converge to a value described by the diagonal ensemble, $O_{\rm DE}$:
\begin{align}
 \langle O(t) \rangle &= \sum_{\alpha,\beta} c_\beta^* c_\alpha e^{-i(E_\alpha-E_\beta)t} \bra{\beta}O\ket{\alpha} 
 \nonumber \\
 & \rightarrow \sum_\alpha |c_\alpha|^2 \bra{\alpha} O \ket{\alpha} \equiv O_{\rm DE}.
\end{align}
In the limit on the right-hand side, any off-diagonal contributions $\bra{\alpha} O \ket{\beta}$ have been averaged away by the oscillatory factor.

In contrast to the diagonal ensemble, which may keep some memory of the initial state, thermal ensembles define the state of a system only by a few thermodynamic parameters which are fixed by the initial conditions. For example, if the canonical ensemble is applied, the state of the system is defined by the temperature $T$. On the other hand, to associate a temperature with the state of the system we shall evaluate the ensemble average $E_{\rm th}(T)$ of the energy:
\begin{align}
\label{ET}
 E_{\rm th}(T) = \frac{1}{Z} {\rm Tr}\left( H e^{-H/T} \right),
\end{align}
where the Boltzmann constant has been set to 1, and $Z$ denotes the partition function $Z={\rm Tr} e^{-H/T}$.
Comparison of the thermal energy $E_{\rm th}(T)$ with the system energy, $E= \bra{\Psi_{\rm ini}} H \ket{\Psi_{\rm ini}}$, yields an effective temperature. With this, one is able to calculate thermal expectation values of any observables:
\begin{align}
\label{CE}
 O_{\rm CE}(T) = \frac{1}{Z} {\rm Tr}\left( O e^{-H/T} \right).
\end{align}

The observable which, in the present study, will be used to track the system is the harmonic oscillator level of the impurity. Experimentally, this quantity can be determined by reducing the height of the trap. At a certain height, the impurity escapes the trap, and from this value, the oscillator level of the impurity can be determined. Formally, the corresponding operator reads $m_{\rm imp} = \sum_m m \ket{n_m}_\downarrow\bra{n_m}_\downarrow$. It acts only on the impurity (that is the $\downarrow$-component), and counts the number of particles $n_m$ in each orbital $m$, weighted by the level number. We can associate $\langle m_{\rm imp} \rangle$ with the energy of the impurity.

\section{Static properties}
\label{Sec:Stat}
Before studying the relaxation dynamics of the system in the next section, let us first discuss some static properties. As has been demonstrated in Ref. \cite{jochim13}, it is possible to measure the interaction energy of the impurity with high accuracy. The measurement is based on determining the resonance frequency for an internal transition of the impurity atom, which, in the presence of majority atoms, is shifted by the amount of the interaction energy. Increasing the number of majority atoms from one to five, it was shown that the interaction energy quickly approaches the theoretical predictions made for an impurity in a homogeneous Fermi sea \cite{mcguire}. 
Here, we will equip the majority atoms with a dipole moment, that is, the Fermi ``sea'' becomes interacting. As we show below, the interaction energy of the impurity provides information also about the strength of the dipolar interactions, although they are restricted to the majority atoms.

\begin{figure}[t]
\centering
\includegraphics[width=0.45\textwidth, angle=0]{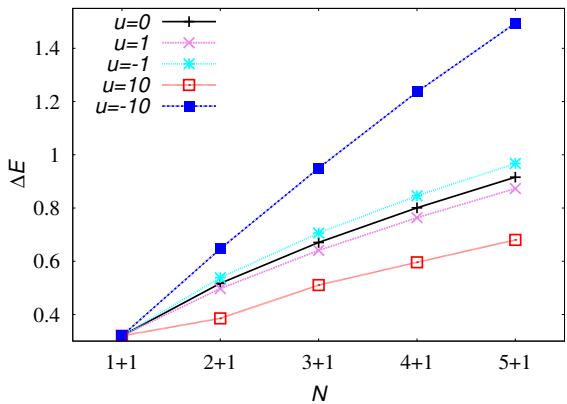} 
\caption{\label{fig1} (Color online)
Interaction energy of an impurity atom with several majority atoms as a function of the system size. The black line shows the behavior for a system of non-interacting majority atoms, $u=0$. Repulsive dipolar interactions between the majority atoms, $u>0$, increase the interaction energy, whereas attractive dipolar interactions decrease the interaction energy.
}
\end{figure}

The system is described by the Hamiltonian $H$ of Eq. (\ref{H}). Two control parameters, $u$ and $g$, allow to adjust independently the interaction between the majority atoms, and the interactions with the impurity. Using numerical diagonalization we have determined the total energy $E(u,g)$ for different dipolar and contact interaction strengths. We obtain the interaction energy of the impurity with the majority atoms as the difference $E(u,g)-E(u,0)$. As shown in Fig. \ref{fig1}, repulsive interactions between the majority atoms ($u>0$) lead to a reduced interaction energy of the impurity, while attractive dipolar interactions ($u<0$) will increase the impurity's interaction energy. 

An explanation for this behavior are density changes due to the dipolar interactions: Repulsive interactions reduce the local density of majority atoms near the trap center. Clearly, this also leads to decreased spatial overlap between majority atoms and impurity, since the impurity predominantly occupies the lowest oscillator level, and thus has an amplitude maximum at the center. As a consequence, the contact interaction energy between both species is reduced. Oppositely, attractive dipolar interactions lead to an increased density of majority atoms  in the center of the trap, and therefore enhance the contact interactions.

\section{Dynamic properties - Case Study}
\label{Sec:Dyn}
We will now investigate how the dipolar interactions between the majority atoms modify the properties of the system dynamics. Again, the impurity atom provides a tool for controlling and measuring the system. More concretely, we propose to study the time evolution of a system out of equilibrium by tracking the energy of the impurity. Numerically, determining the  time evolution is an extremely hard task, as knowledge of all eigenstates is required, while exact diagonalization algorithms work most efficiently if only few eigenvectors are obtained. We therefore restrict the system size to 3+1 atoms. Taking into account parity symmetry of the Hamiltonian, and restricting the single-particle basis to 22 (20) states, we have Hilbert space blocks of 16940 (11400) states. We have checked our results for convergence with respect to increases in the single-particle basis. Clearly, truncation errors are typically more pronounced for excited states than for the ground state, and time evolution exponentiates every error. Accordingly, in some cases, convergence can be achieved only on short time scales.


\begin{figure}[t]
\centering
\includegraphics[width=0.45\textwidth, angle=0]{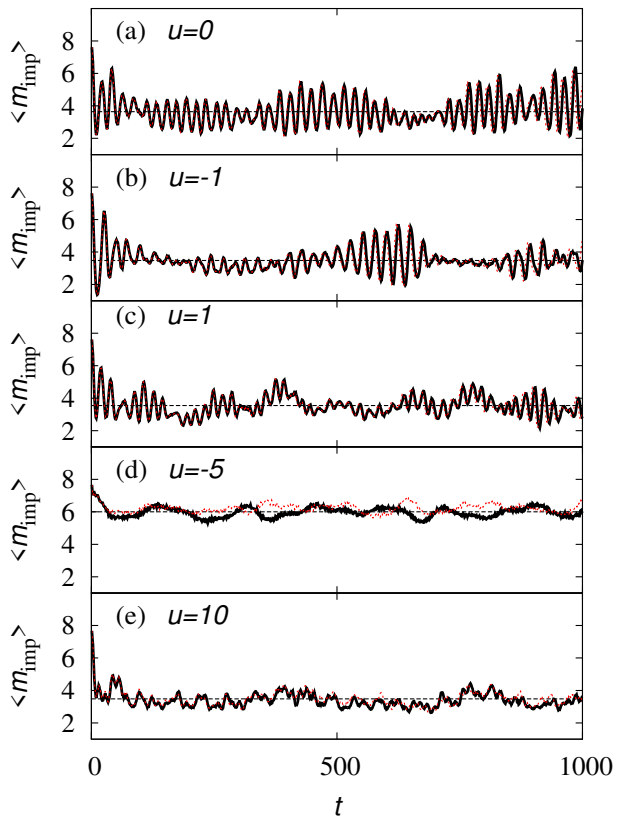} 
\caption{\label{fig2} (Color online) 
Time evolution tracked by the average level number $\langle m \rangle_{\rm imp}$ of the impurity. The initial state is a Fock state with the three majority atoms in the three lowest levels, and the impurity occupying the 8th excited level, $n_{\rm imp}=8$. In all plots, the Hamiltonian $H$ contains a repulsive contact interaction of the majority atoms with the impurity ($g=1$). The strength $u$ of the dipolar interactions between the majority atoms takes the value denoted in each plot. The horizontal dashed lines show the corresponding diagonal ensemble average. The red dotted line indicates whether the numerical treatment is converged: The black line shows the results taking into account 20 (a-c) or 22 (d,e) harmonic oscillator levels. The red dotted shows the results for a smaller Hilbert space, taking into account 18 (a-c) or 20 (d,e) levels.}
\end{figure}

\begin{figure}[t]
\centering
\includegraphics[width=0.45\textwidth, angle=0]{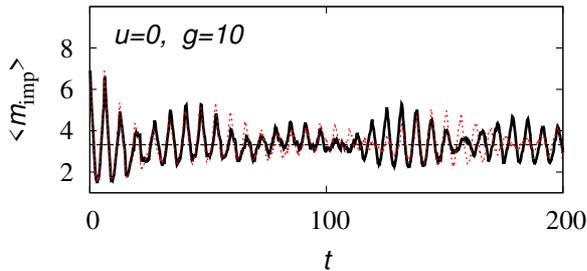} 
\caption{\label{fig3} (Color online) 
Time evolution as in Fig. \ref{fig2}, but with strong contact interactions, $g=10$, and no dipolar interactions, $u=10$. The black line (red dotted line) shows the results taking into account 22 (20) harmonic oscillator levels. The horizontal dashed line indicates the diagonal ensemble average.  }
\end{figure}

We choose an easy-to-prepare initial state at relatively low energy. A convenient choice, taken for Fig. \ref{fig2}, is a Fock state with the three majority atoms populating the three lowest levels. In the absence of an impurity, this is the ground state of a system without dipolar interactions. The impurity is used to bring the system out of equilibrium. To have pronounced effects, the impurity shall occupy a level significantly above the equilibrium value. We have chosen $m_{\rm imp} = 8$ in Fig. \ref{fig2}. 
We then studied the time evolution of this initial state under the Hamiltonian $H$ of Eq. (\ref{H}) for a contact interaction strength $g=1$, and for different values $u$ for the dipolar interactions between the majority atoms. Being a Fock state, the initial state is an (excited) eigenstate of the non-interacting Hamiltonian, $g=u=0$. This allows to view the scenario also as an interaction quench, during which both contact and dipolar interactions are suddenly switched on.

Even without quantitative analysis, it is obvious from Fig. \ref{fig2} that relaxation does not take place in the absence of dipolar interactions [Fig. \ref{fig2}(a)]. In this case, the system evolution is characterized by an oscillatory behavior for all times. In particular, even after a long relaxation time, the amplitude $\delta \langle m_{\rm imp} \rangle$ of these oscillations still attains large values, $\delta \langle m_{\rm imp} \rangle > 1$. On average for long times ($350<t<1000$), the impurity level $\langle m_{\rm imp} \rangle(t)$ oscillates around a mean $\mu=3.7$, close to the diagonal ensemble average of 3.6, with a standard deviation $\sigma=0.9$. 

One might argue that the observed regularity of the dynamics is due to the weakness of interactions, and therefore the proximity of the initial state to an eigenstate of the Hamiltonian $H$. 
As a quantitative measure for the amount of Hamiltonian eigenstates participating in the initial state, we define a kind of entropy
\begin{align}
 S = - \sum_\alpha |c_\alpha|^2 {\rm ln} |c_\alpha|^2,
\end{align}
with the $c_\alpha$ the coefficients of a decomposition of the initial state in the eigenbasis of $H$, cf. Eq. (\ref{Psi}). If $S=0$, the initial state is an eigenstate of $H$. If $S={\rm ln} D$, $D$ eigenstates contribute with equal weights. Indeed, for the Hamiltonian of Fig. \ref{fig2} (a), the entropy of the initial state takes only a moderate value, $S=2.7$. Also, the average interaction energy $\langle V \rangle = 0.2$ is small compared to the total energy, $\langle H \rangle = 11.2$. 
However, even for an increased contact interaction $g=10$ (and $u=0$), the dynamics remains regular, see Fig. \ref{fig3}, although now entropy and interaction energy are significantly larger, $S=4.1$ and $\langle V \rangle = 2.4$. This suggests that it is not the weakness of the quench which led to the regular dynamics in Fig. \ref{fig2} (a). We note that for $g=10$, the time evolution is quantitatively not converged for large times $t \gtrsim 100$. Qualitatively, however, the evolution remains regular on all time scales, independently from the number of single-particle states taken into account.

Next, we consider the case of weak dipolar interactions between the majority atoms, in addition to weak contact interactions with the impurity, see Fig. \ref{fig2} (b,c). The oscillation amplitudes are reduced compared to the case without dipolar interactions, with a long-time mean $\mu = 3.5 \pm 0.7$ for attractive interactions ($u=-1$), and $\mu=3.6 \pm 0.5$ for repulsive interactions ($u=1$), in agreement with the diagonal ensemble average (3.5 in both cases). However, in particular for the attractive system, periods of strong oscillations occur repeatedly even for long relaxation times, keeping the system away from a steady-state. In both the attractive and the repulsive case, the entropy of the initial state is $S=2.8$. Additionally to the contact interaction energy, $\langle V_{\rm c} \rangle= 0.2$, the total interaction energy now contains also a dipolar contribution: $\langle V \rangle = \langle V_{\rm dd} \rangle + \langle V_{\rm c} \rangle$. The dipolar contribution takes the values $\langle V_{\rm dd} \rangle =  \pm 0.4$ for $u = \mp 1$.

In Fig. \ref{fig2} (d,e), we turn our attention to strong dipolar interactions, $u=-5$ and $u=10$. Since attractive interactions have a stronger tendency to populate higher oscillator levels, we are not able to achieve numerical results which remain convergent for long times. As seen in Fig. \ref{fig2} (d), the time evolution for $u=-5$ is quantitatively not converged for $t \gtrsim 50$. This means that only the initial decay of $\langle n_{\rm imp}\rangle(t)$ is captured accurately. However, for larger times the evolution consists only of small-valued fluctuations around a mean value corresponding to the diagonal ensemble average ($\mu = 6.0 \pm 0.3$ for a basis with 22 states, and $\mu=6.2 \pm 0.3$ for a basis with 20 states). This suggests that the system quickly reaches a steady state, despite the remarkably small entropy of the initial state, $S=2.4$. The interaction energy is $\langle V_{\rm dd} \rangle = -2.1$. 

For strongly repulsive dipolar interactions, shown in Fig. \ref{fig2} (e), we are able to obtain fairly well converged results even for long relaxation times. The system quickly evolves towards the average of the diagonal ensemble (3.4), around which it fluctuates in an erratic manner with a standard deviation $\sigma=0.4$. This means that the deviations from the average are only slightly smaller than for weak repulsive interaction, Fig. \ref{fig2} (c). Instead, it is the absence of oscillatory behavior which make the strongly interacting system appear significantly more relaxed than the weakly interacting one. Both, average dipolar interaction energy and entropy of the initial state are large: $\langle V_{\rm dd} \rangle = 4.1$ and $S=4.4$.

\begin{table}
\begin{tabular}{|l|c|c|c|}
\hline
 & $\mu \pm \sigma$ & $S$ & $\langle V_{\rm dd} \rangle $ \\
\hline 
 $g=1, u=0$ \ \ &  $3.7\pm0.9$ &2.7 & 0 \\
\hline
 $g=1, u=-1$ \ \ & $3.5\pm0.7$ & 2.8 & -0.4 \\
\hline
 $g=1, u=1$ \ \ & $3.6\pm0.5$ & 2.8 & 0.4 \\
\hline
 $g=1, u=-5$ \ \ & $6.0\pm0.3$ & 2.4 & -2.1 \\
\hline
 $g=1, u=10$ \  \ & $3.4\pm0.4$ & 4.1 & 4.1 \\
\hline
 \end{tabular}
\caption{Quantitative analysis of the data presented in Fig. \ref{fig2}.}
\end{table}

\begin{figure}[t]
\centering
\includegraphics[width=0.45\textwidth, angle=0]{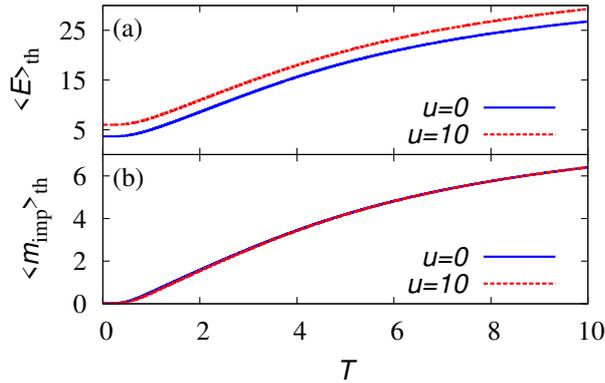} 
\caption{\label{fig32}
(Color online) Canonical ensemble averages of (a) energy $E_{\rm th}$ and (b) impurity oscillator level $\langle m_{\rm imp} \rangle$, as a function of temperature $T$.}
\end{figure}

Finally, we may ask whether this relaxed state is also thermalized. In the canonical ensemble, energy $E$ is associated with temperature $T$ according to Eq. (\ref{ET}). This leads to a gauge curve $E(T)$ plotted in Fig. \ref{fig32} (a) for $g=1$, and $u=0$ and 10. From this curve, we infer an effective temperature $T=3.2$ for the system with strong dipolar interactions. On the other hand, using the canonical ensemble Eq. (\ref{CE}), we can also evaluate the thermal average of $n_{\rm imp}$ for any given temperature. The question is whether $n_{\rm imp}$ relaxes to the value which corresponds to the effective temperature defined via the energy. The temperature dependence of $n_{\rm imp}$, which has basically no dependence on $u$, is plotted in Fig. \ref{fig32} (b). For $T=3.2$, we find the thermal average $\langle n_{\rm imp} \rangle_{\rm th} \approx 2.8$, which is significantly below the temporal average in Fig. \ref{fig2} (e), around 3.5. This comparison shows that the system, although it relaxes, does not thermalize.

\section{Dynamic properties - Discussion}
\label{Discussion}
Let us summarize the main findings of the case studies presented in the previous section:
\begin{itemize}
 \item Relaxation towards the diagonal ensemble average occurs in the presence of sufficiently strong dipolar interactions between the majority atoms.
 \item The system dynamics remains oscillatory for a system with purely contact interactions. This seems not to be related to a smaller interaction energy, nor to a smaller participation entropy.
 \item The system, even when it relaxes, does not thermalize.
\end{itemize}

\begin{figure}[t]
\centering
\includegraphics[width=0.45\textwidth, angle=0]{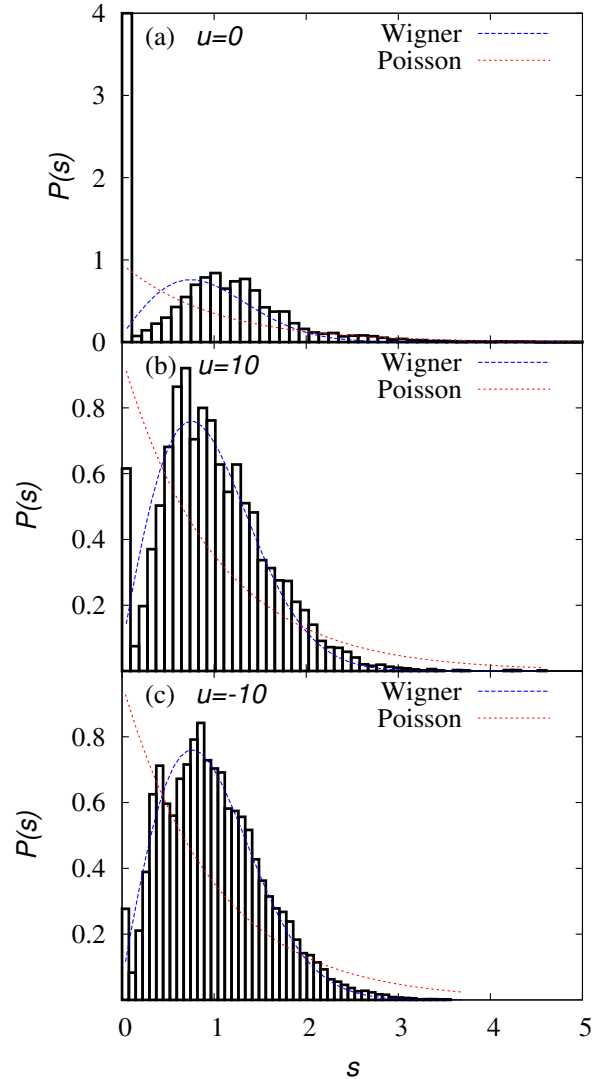} 
\caption{\label{fig7} (Color online)
Unfolded level spacing in energy spectra of $H$ with $g=1$ and for different $u$. For normalizing the distribution, we did not take into account levels at $s=0$.
}
\end{figure}

Let us now try to understand the differences in the relaxation behavior from a more general point of view. A concept which will prove useful is the distinction between chaotic and regular dynamics: While regular dynamics preserves the coherences of the initial state, chaotic evolution should quickly lead to relaxation. Then, the different behavior should be reflected by the energy spectra. In quantum mechanics, chaotic behavior is characterized by level repulsion. For a system with time-reversal symmetry, the Hamiltonian matrix is real-symmetric, and random matrix theory, cf. Ref. \cite{haake-book}, tells us that the level spacing should follow a Wigner distribution function, $P_{\rm Wigner}(s) = \frac{\pi}{2} s e^{-s^2 \pi/4}$. Clearly, the maximum  of this distribution is located at finite spacings $s=\sqrt{2/\pi}$. On the other hand, such level repulsion is not expected for a system with regular dynamics. In this case, conserved quantities will allow energy levels to cross, that is, correlations between the levels are absent. This may lead to a Poissonian level spacing distribution, $P_{\rm Poisson}(s) = e^{-s}$, with a maximum for $s=0$.

Before analyzing the level spacing distribution, it is necessary to unfold the energy spectra. Naturally, level spacings will be larger in spectral regions with a low density of states than in dense regions, but this obviously has nothing to do with correlations between energy levels. By unfolding the energy spectrum, we shall guarantee an overall homogeneous density of states, such that level correlations become the only source for variations in the level spacing. To that aim, we follow the unfolding procedure described in Ref. \cite{haake-book}. The first step is to smooth the discrete density of states via a convolution with a Gaussian:
\begin{align}
 \rho_{\rm smooth}(E) \frac{1}{N} \sum_{i=1}^N \frac{1}{\Delta \sqrt{\pi}} e^{-(E-E_i)^2/\Delta^2}.
\end{align}
We have chosen the width $\Delta$ of the Gaussian to be given by five times the mean level spacing. The smoothened density of states also leads to a smooth staircase function $\Sigma_{\rm smooth}(E)$, that is, a smooth cumulated density of states:
\begin{align}
 \Sigma_{\rm smooth}(E)= \int_{-\infty}^{E} dE' \ \rho_{\rm smooth}(E')
\end{align}
The unfolded energy levels $e_i$ are then associated with $e_i= N \Sigma_{\rm smooth}(E_i)$, and the unfolded level spacings are $s_i=e_i-e_{i-1}$.

We have evaluated the unfolded level spacing distribution for the energy spectrum of $H$ with $g=1$ and $u=0, 10, -10$, for 3+1 atoms, taking into account 20 harmonic oscillator levels. 
In the presence of dipolar interactions, as seen in Fig. \ref{fig7} (b,c), the levels spacing is well described by the Wigner distribution, except for a somewhat increased number of levels near $s=0$ in the case of repulsive interactions, see Fig. \ref{fig7} (b). Nevertheless, the distributions clearly indicate level repulsion, and thus chaotic dynamics. This is in agreement with the relaxation properties discussed in the previous section. More complicated to interpret is the distribution in the absence of dipolar interactions, Fig. \ref{fig7} (a). Neither the Wigner nor the Poisson distribution capture the level spacing distribution. On the other hand, an eye-catching property of the distribution is the exorbitantly large number of levels near $s\approx 0$, with around 4400 of 11399 spacings peaked around zero. This number is not compatible even with a Poisson distribution, for which around 10\% of the levels should be located within $s<0.1$ \footnote{In Fig. \ref{fig7} (a), only levels at non-zero $s$ have been taken into account for normalizing the distribution.}.
The distribution of the remaining levels is neither described by a Poisson nor by a Wigner function, even if we exclude the $s=0$ levels from the normalization (which leads to a normalized distribution of the remaining levels). The Wigner distribution catches well only the behavior at intermediate spacing, $1 \lesssim s \lesssim 2$, whereas the tail of the distribution has Poissonian shape. It might be that the huge number of degenerate levels spoils our unfolding procedure, and/or that a mixed phase space, consisting of both chaotic and regular regimes, leads to such unconventional distribution function. As a conclusion, the dominant feature of the distribution are the abundantly many degeneracies, and therefore we may take the observed level spacing as an indicator for a regular system dynamics.

An observation which might, at least partly, explain the large number of nearby levels is the ``quasi''-integrability of the model without dipolar interactions: First, there is an exact solution for 1+1 particles in a harmonic trap \cite{busch}, and a quasi-exact ansatz for $N+1$ particle \cite{levinsen}. Second, in the homogeneous case with $N+1$ particles, the system is integrable via Bethe ansatz \cite{mcguire,guan-review}. Finally, although no exact ground state solution is known for the trapped system with $N+1$ fermions, exact wave functions are easily obtained for several excited states due to the so-called ``fermionization'' of the problem.  In fact, the degeneracies in the energy spectra can be traced back to these fermionized solutions. The idea behind the fermionization is to fill $N+1$ different harmonic oscillator levels in a fermionic way, that is, by assuming the wave function to be given by the corresponding Slater determinant. Such spatial wave function is fully antisymmetric, irrespective of the particle spin. Obviously, it completely suppresses the contact interactions, and in the absence of dipolar interactions, it provides an eigenstate with the energy given by the single-particle energy of the occupied harmonic oscillator levels. 
For larger energies, there is a rapidly growing number of ways how $N+1$ particles can be distributed, leading to a huge number of degenerate levels. In this context,  it is interesting to note that for strongly repulsive contact interactions, the energies of all eigenstates approach the values given by these fermionized solutions \cite{sowinski13,levinsen}. In that case, energy levels are either degenerate, or separated by the integer spacing $\hbar \omega$, which apparently would lead to a very untypical bimodal level spacing distribution.

\begin{figure}[t]
\centering
\includegraphics[width=0.45\textwidth, angle=0]{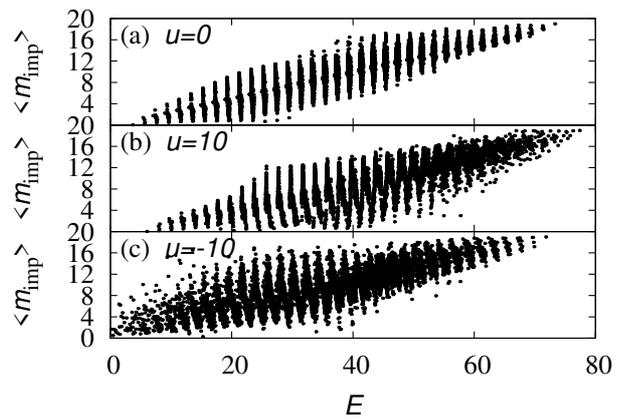} 
\caption{\label{fig8}
We plot the average occupation $m_{\rm imp}$ vs. the energy for all eigenstates of $H$, with $g=1$ and for different $u$.
}
\end{figure}

Let us finally study why the system, even for cases in which it equilibrates, lacks thermalization. A hypothesis explaining thermalization of an isolated quantum system is the eigenstate thermalization hypothesis. It assumes that, for any eigenstate, the quantum average of an operator and the corresponding eigenenergy are correlated, that is, all eigenstates in the vicinity of some energy $E$ have the same expectation values with respect to relevant observables $O$. Moreover, these values are assumed to be ``thermalized'', that is they shall coincide with the thermodynamic average. We test the eigenstate thermalization hypothesis by plotting the average occupation $\langle m_{\rm imp} \rangle $ versus the energy for all eigenstates in Fig. \ref{fig8}. In both cases, with or without dipolar interactions, the values of $\langle m_{\rm imp} \rangle$ are spread over a broad range for almost any energy in the spectrum. Accordingly, the eigenstate thermalization hypothesis does not hold, and fixing the energy in a microcanonical ensemble will not fix the expectation value of $m_{\rm imp}$. This explains why thermalization does not take place.

\section{Summary and Outlook}
In summary, we have suggested to probe the dynamics of a Fermi system trapped in 1D by exciting an impurity and tracking the oscillator level of the impurity. We have studied a system with dipolar interactions, and have contrasted its dynamics to the dynamics of a system with purely contact interactions. While the latter shows quantum collapse and revival effects even after long relaxation times, the system with sufficiently strong dipolar interactions relaxes quickly to its diagonal ensemble average. We relate this finding to the level spacing distribution which indicates chaotic behavior in the presence of dipolar interactions. In the case without dipolar interactions, a huge number of degenerate levels characterizes the level spacing distribution.

Even despite the small system size of only four fermions, we encounter cases where the time evolution cannot be computed faithfully on long time scales. This demonstrates the need for better computational techniques, and/or alternative approaches such as quantum simulations. Our work suggests to explore dynamical aspects in the experimental setting of Ref. \cite{jochim13}. In view of the recent experimental progress with dipolar atoms \cite{ferlaino,adams15,laburthe15} and molecules \cite{debbie13}, scenarios as the one studied in this paper are becoming experimentally feasible.

\acknowledgments
I am grateful to Maciej Lewenstein for reading and commenting the manuscript and for discussions. I also wish to thank Miguel Angel Garcia-March, Tomek Grining, Pietro Massignan, and Michal Tomza for discussions. Financial support from EU grants OSYRIS (ERC-2013-AdG Grant No. 339106), SIQS (FP7-ICT-2011-9 No. 600645), QUIC (H2020-FETPROACT-2014 No. 641122), EQuaM
(FP7/2007-2013 Grant No. 323714), Spanish Ministry grant FOQUS (FIS2013-46768-P), and Fundaci\'o Cellex is acknowledged.

\appendix
\section{Dipole matrix elements}
We need to evaluate matrix elements $ V_{ijkl}^{\rm dipolar} = \int_{-\infty}^{\infty} dx \ dx' \ U(x/\lambda-x'/\lambda) \varphi_i(x) \varphi_j(x') \varphi_k(x) \varphi_l(x')$ for a two-body potential
\begin{align}
 U(x)=-2|x|+\sqrt{2\pi}(1+x^2)e^{x^2/2}{\rm erfc}(x/\sqrt{2}).
\end{align}
The orbitals $\varphi_i(x)$ are the usual harmonic oscillator levels
\begin{align}
 \varphi_j(x) = \sqrt{\frac{1}{\pi^{1/2} 2^j j!}} e^{-x^2/2} H_n(x) \equiv N_j e^{-x^2/2} H_n(x),
\end{align}
where $H_n(x)$ are the Hermite polynomials, and $N_j$ denotes the normalization constant of the wave function. To evaluate the integral, we replace the coordinates $x$ and $x'$ by relative coordinates $r=x-x'$ and center-of-mass coordinates $R=x+x'$. The integral over the center-of-mass part, $I_R^{(i)}$, reduces to
\begin{align}
 I_R^{(i)} \equiv  \int_{-\infty}^{\infty} R^{i-1} e^{-R^2/2} dR = 2^{i/2-1} \left[1+(-1)^{i+1}\right] \Gamma\left(\frac{i}{2}\right),
\end{align}
with $i$ some positive integer. Also the integral over the relative coordinate, $I_r^{(i)}$, turns out to have a compact solution
\begin{widetext}
\begin{align}
  I_r^{(i)} \equiv& \int_{-\infty}^{\infty} r^{i-1} e^{-r^2/2} \left[-2 \Big|\frac{r}{\lambda}\Big| \sqrt{2\pi}(1+\frac{r^2}{\lambda^2})e^{(r/\lambda)^2/2}{\rm erfc}\left(\Big| \frac{r}{\sqrt{2}\lambda} \Big| \right) \right] dr = 
 \frac{1}{\lambda} 2^{-\frac{1}{2}-\frac{i}{2}} \left[-1+(-1)^i\right] \times
   \nonumber \\ & \times
 \left[ 2^{i+1} \Gamma\left(\frac{i+1}{2} \right) - 2\sqrt{\pi} \lambda^{i+1} \Gamma(i) \ _2 \tilde F_1\left(\frac{i}{2},\frac{i+1}{2},\frac{i+2}{2};1-\lambda^2\right) - \sqrt{\pi} \lambda^{i+1}\Gamma(i+2) \ _2 \tilde F_1\left(\frac{i}{2},\frac{i+1}{2},\frac{i+2}{2};1-\lambda^2\right) \right] ,
\end{align}
where $_2 \tilde F_1(a,b,c;z)$ is the regularized hypergeometrical function.
\end{widetext}

Any matrix element $V_{ijkl}^{\rm dipolar}$ can be decomposed into a sum over products of these two integrals $I_R^{(i)}$ and $I_r^{(i)}$, with prefactors stemming from the Hermite polynomials and the normalization factors $N_j$ of the orbitals. However, care must be taken when numerically evaluating the matrix element in high orbitals. Since the numeric values of the integrals become large, while the normalization factors become small, it is crucial to keep symbolic expressions as long as possible to avoid numeric errors. A piece of Mathematica code which performs the decomposition and evaluates the matrix element $V_{n1n2n3n4}^{\rm dipolar}$ reads
\begin{widetext}
\begin{verbatim}
integral[n1_, n2_, n3_, n4_,Lambda_] := Module[{List1, List2, IR, Ir, r, R},
  List1 = CoefficientList[HermiteH[n1,(r+R)/2]HermiteH[n2,(R-r)/2]
          HermiteH[n3,(r+R)/2]HermiteH[n4,(R-r)/2],R];
  IR = Table[I_R[i], {i, 1, Length[List1]}];
  List2 = CoefficientList[Dot[List1, IR], r];
  Ir = Table[I_r[i,Lambda], {i, 1, Length[List1]}];
  N[Dot[List2, Ir] norm[n1] norm[n2] norm[n3] norm[n4]]]]
\end{verbatim}
where \verb I_R[i]  is given by $I_R{(i)}$, \verb I_r[i,lambda]  is given by $I_r{(i)}$ for some choice of $\lambda$, and \verb norm[n1] is the normalization factor $N_{n1}$.
\end{widetext}


\end{document}